\documentclass{article}

\usepackage[dvips]{graphicx}
\usepackage{amssymb,amsfonts,amsmath}
\usepackage{textcomp}

\usepackage{color}
\usepackage[bf]{caption}

\begin{document}

\title{Collective dynamics of active cytoskeletal networks}

\author{Simone K\"{o}hler, Volker Schaller and Andreas R. Bausch \\
\vspace*{0.5em}
Technische Universit\"{a}t M\"{u}nchen, Garching, Germany}

\date{}

\maketitle


\textbf{
Self organization mechanisms are essential for the cytoskeleton to adapt to the requirements of living cells. They rely on the intricate interplay of cytoskeletal filaments, crosslinking proteins and molecular motors. Here we present an \emph{in vitro} minimal model system consisting of actin filaments, fascin and myosin-II filaments exhibiting pulsative collective long range dynamics. The reorganizations in the highly dynamic steady state of the active gel are characterized by alternating periods of runs and stalls resulting in a superdiffusive dynamics of the network's constituents. They are dominated by the complex competition of crosslinking molecules and motor filaments in the network: Collective dynamics are only observed if the relative strength of the binding of myosin-II filaments to the actin network allows exerting high enough forces to unbind actin/fascin crosslinks. The feedback between structure formation and dynamics can be resolved by combining these experiments with phenomenological simulations based on simple interaction rules.
}

\paragraph{Introduction}
The cytoskeleton of eukaryotic cells is a highly flexible and adaptable scaffold that undergoes constant remodeling to meet their changing needs. With its unique static and dynamic properties it facilitates tasks as complex and diverse as cell division, cell locomotion or phagocytosis. Intracellular patterns can emerge, as has been observed in the apical constriction during Drosophila gastrulation, where the tissue rearrangement is driven by pulsed contractions of the actin myosin cytoskeleton \cite{Martin:2009p16697,Martin:2010p16699}. Similar self organization mechanisms are of outmost importance in many aspects of cellular development \cite{Karsenti:2008p26467}.
All these processes rely on the intricate interplay between three major components: actin filaments, molecular motors and crosslinking proteins. While a polymer network consisting of filaments and crosslinkers result in a viscoelastic physical gel, molecular motors exert local forces and turn it into an active gel \cite{Joanny:2007p26677}.
The dynamics in these active actomyosin gels can be coordinated in time and space as has been observed for the pulsed constrictions during dorsal closure giving rise to a collective behavior \emph{in vivo} \cite{Solon:2009p43479}.

A processive force exertion of the molecular motors inside a purely viscous environment would lead to ballistic motion of the cytoskeletal material, only limited by the viscous friction of the environment. Inside of a homogenous viscoelastic medium additional relaxation processes occur, which would mainly result in a slowdown of the dynamics.

Yet, the cytoskeleton is far from being homogeneous.
Here, the coordinated action of molecular motors and crosslinking proteins leads to the formation and stabilization of a wide variety of structures that in turn affect the large scale dynamics of the system.
It is the feedback between structure formation and dynamic properties that sets the functional properties of cytoskeletal systems and active gels in general. However, due to the lack of adequate model systems, it remains difficult to address these feedback mechanisms.

Force dissipation in the viscoelastic environment and the inherent heterogeneity should modulate the dynamics in a nontrivial way:
The transport by molecular motors relies on the interaction with discrete and anisotropic filaments, which could suggest the occurrence of anomalous diffusion processes. At the same time, the force exertion inside of crosslinked cytoskeletal networks are expected to induce a force rate dependent rupturing of the crosslinking points \cite{Evans:1997p27714} and thus the locally induced changes of the elastic environment will also affect and modulate the transport and structure formation dynamics. Thereby, it would be conceivable that the connectivity of the filamentous network enables the coordination of the transport and reorganization processes resulting in the appearance of collective modes.

To shed light on the principles underlying the physics of such active gels, to examine their microscopic dynamics and to classify the thereby resulting dynamic structures, we study a reconstituted actin network that is actively set under stress by molecular motors \cite{Koehler:2011}. While crosslinking proteins are required for the mechanical stability of the actin network, molecular motors introduce an active component to the networks \cite{Humphrey:2002p25122,Smith:2007p19085,Mizuno:2007p2117,Koenderink:2009p2973}. The presence of only motor filaments and ATP does not suffice to induce any reorganization or structure formation in an actin solution. Only the combination of motors and crosslinking molecules can result in structure formation \cite{Backouche:2006p537} and even can induce a macroscopic contraction of cytoskeletal \emph{in vitro} networks \cite{Kane:1983p5461,Janson:1991p18885,Bendix:2008p5329}.

Here we show that in an active and crosslinked network pulsative collective modes develop, which depend critically on the interaction strength of the myosin-II motor filaments. Our recently introduced minimal approach \cite{Koehler:2011} using a reconstituted actin/fascin network in presence of myosin-II filaments enables a backtracking of the pulsative collective dynamic properties to the structure formation: The connectivity of the network is the basis for the coordinated reorganization of the cytoskeletal structure in the steady state. The dynamics are dominated by the complex competition of crosslinking molecules and motor filaments in the network. The key player is the relative strength of the binding of myosin-II filaments to the actin network: only a high binding affinity allows exerting high enough forces to unbind actin/fascin crosslinks resulting in network activity. The resulting dynamics can be described by an anomalous diffusion of the network's constituents.

\begin{figure}[h!]
  \centering
    \includegraphics[width=.8\textwidth]{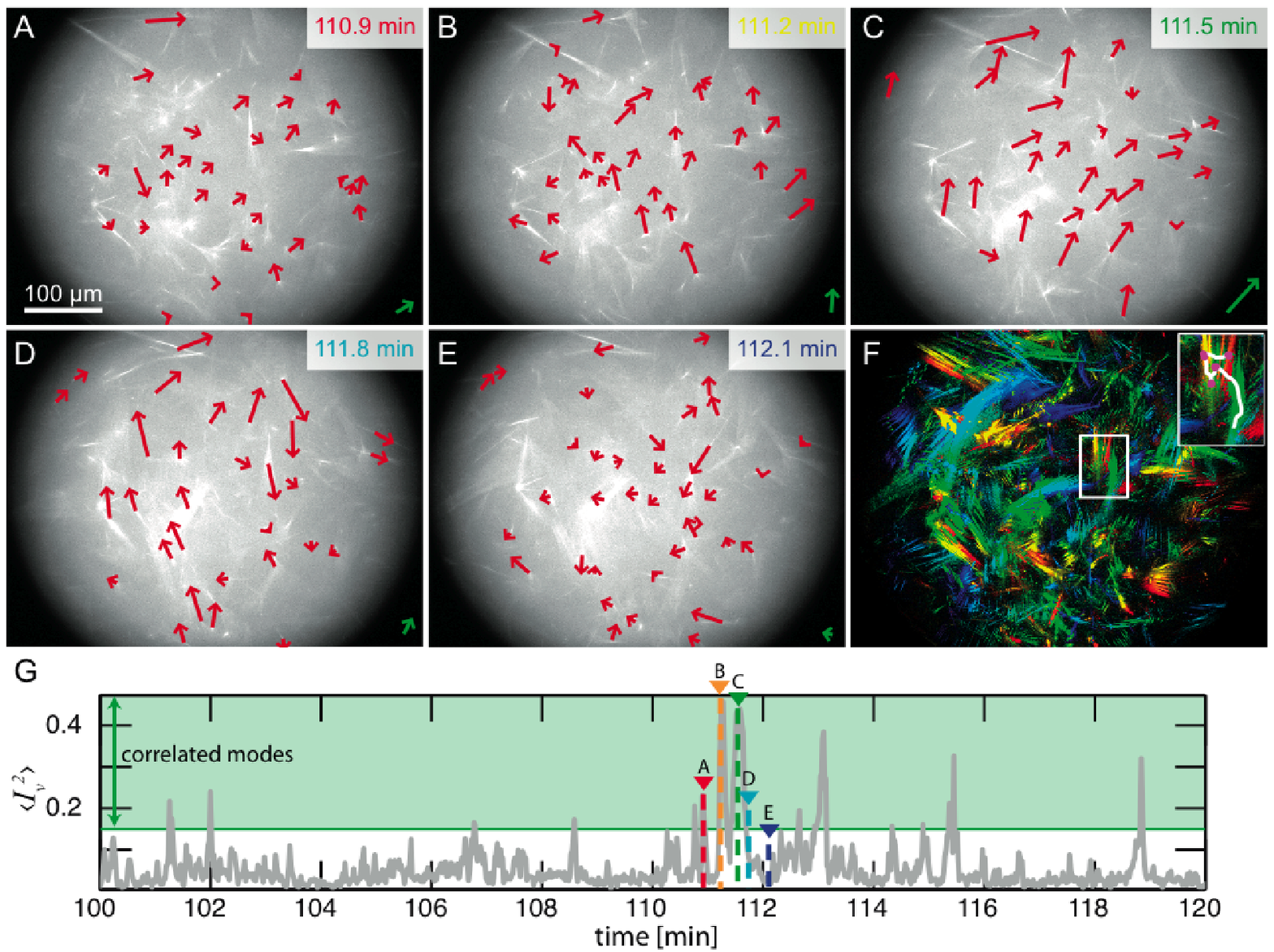}
     \caption{\textbf{Structure and dynamics of actin/fascin/myosin networks.}
       Fluorescence micrographs (\textbf{A--E}) of active actin networks at indicated times after initiation of polymerization (0.1\,mM ATP) show the dynamic reorganization within the network. Red arrows indicate the movement of individual points in the network with lengths 20-fold magnified and a time-average over three frames. Green arrows show the resulting overall movement in the field of view (lengths are 40-fold magnified).  These long range reorganization processes are summarized in the colored time overlay (red to blue, \textbf{F}). Due to the low magnification used in this experiment, very small structures cannot be resolved. Thus, the connectivity between the structures is higher than can be seen in these fluorescence micrographs. The trajectory of an individual structure exhibits persistent runs (inset, white line) intermitted by stall periods (inset, magenta dots). These run and stall phases are not only observed for individual structures, but also the whole network in the field of view exhibits pulsatile collective dynamics with movements being coordinated in time and direction: In stall phases, individual structures move predominantly for only short periods and in random directions (\textbf{A,E}). The stall phases are followed by periods with high activity in the entire network. During these run phases, the better part of the network shows long, persistent runs (\textbf{B,C}). The degree of coordination of the movement is measured by the average squared velocity cross-correlation function $\langle I_v^2\rangle$, in which phases of collective movement are reflected as peaks. In such collective phases that last for up to 30 sec, the majority of the identified structures moves in approximately the same direction (\textbf{G}).
}
  \label{fig:traje}
\end{figure}

\paragraph{Results}
Active gels composed of actin filaments, the crosslinker fascin and skeletal muscle myosin-II filaments in presence of ATP undergo drastic structural rearrangements. The dynamics within the network is characterized by a succession of persistent runs and stalling events of individual actin/fascin/myosin-II structures (fig.~\ref{fig:traje}A-F and supplemental movie~S1). 
In general, the individual structures move independently. Yet, for short time intervals of about 30\,s they can also syncronize their movements. These pulsative collective modes are characterized by a coordinated movement in time and space of a few or even up to nearly all visible structures of the network (fig.~\ref{fig:traje}G). Structures which are up to hundreds of microns apart are synchronized in their movement: during run phases their directional motion is coordinated (fig.~\ref{fig:traje}C) and the active movement starts and stops almost simultaneously. This coordinated movement is not altering the steady state of the system and is not linked to a macroscopic contraction, which only occurs at much higher actin concentrations \cite{Koehler:2011}. The coordinated reorganization is only possible if the structures are connected, which is a prerequisite for the observed long range nature of the interaction. Molecular motors can only transport and exert forces between filaments. Thus, it is the connectivity and discrete nature of the network together with the force exertion mechanism, which enables the collective motion to occur. 
Introducing the velocity cross-correlation function $I_v(\mathbf{r})$ collective movements can directly be quantified. For each time point the average of the squared correlation function $\langle I_v(\mathbf{r})^2\rangle_r$ reflects the level of correlation in the system. High values of $\langle I_v(\mathbf{r})^2\rangle_r$ relate to a highly collective movement, while low values close to zero indicate intervals lacking collective modes (fig.~\ref{fig:traje}G and fig.~S1). At 0.1\,mM ATP, for a correlation cutoff $k = 0.15$ about 9\,\% of the frames show correlated movement (see material and methods and movie~S1).

\begin{figure}[h!]
  \centering
  \includegraphics[width=.5\textwidth]{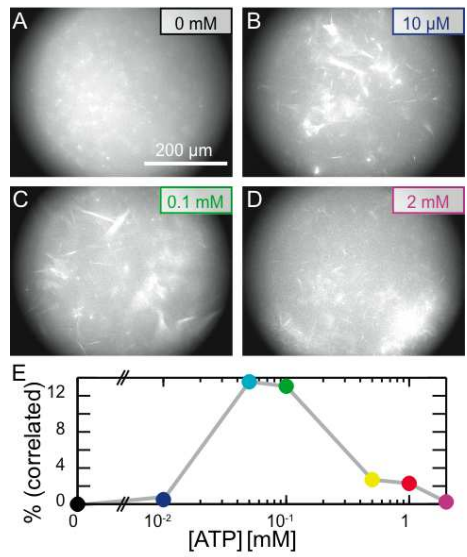}
  \caption{\textbf{Dependence of the dynamics on the ATP concentration. }%
      In abscence of ATP, myosin-II filaments act as nucleation seeds, resulting in a multitude of small, well separated clusters as illustrated in the fluorescence micrograph 90 min after polymerization (\textbf{A}). In prescence of low (textbf{B}), intermediate (textbf{C}) or high (textbf{D}) ATP concentrations, a highly dynamic network of larger structures is formed. The degree of correlated motion determined by the percentage of frames correlated more than 0.15 is dependent on the ATP concentration: Only intermediate ATP concentrations allow for highly correlated movement (\textbf{E}).}
  \label{fig:ATP}
\end{figure}

Microscopically, stall phases are the direct consequence of the presence of the crosslinking molecules while the runs can be ascribed to the activity of the myosin-II filaments. Accordingly, the dynamics and their collective modes should be set by the motor activity. Thus, lowering the motor activity by decreasing the ATP concentration should lead to a continuous decrease of the network dynamics. Yet, the variation of the ATP concentration results in a more complex behavior, as the competition of the motors with the crosslinking molecules comes into play. We observe four distinct regimes (fig.~\ref{fig:ATP}): (i) in the absence of ATP a passive actin/fascin/myosin network consisting of small, diffusing clusters is observed (fig.~\ref{fig:ATP}A and supplemental movie~S2). Their movement is not correlated (fig.~\ref{fig:ATP}E). The passive actin/fascin/myosin network structure is distinct from an actin/fascin bundle network. This can be explained by a competitive binding of fascin and myosin-II to the actin filaments. Moreover, myosin-II filaments seem to act as nucleation seed of the actin polymerization. (ii) At low ATP concentrations (10\,\textmu{}M, fig.~\ref{fig:ATP}B) the low motor activity is limiting the number of runs and the stalls of the network structures are dominating the low dynamics. Consequently, collective modes barely occur in this crosslinker dominated regime. (iii) In the intermediate regime (up to 0.1\,mM, fig.~\ref{fig:ATP}C) the highest correlation is observed. (iv) At high ATP concentrations (fig.~\ref{fig:ATP}D and supplemental movie~S3) correlated motion is again rarely observed. Moreover, at high ATP concentrations the networks become disrupted and huge clusters emerge. The maximal area of these clusters typically exceeds the area of a compact disc with radius of 80\,\textmu{}m whereas structures at lower ATP concentrations remain significantly smaller. The structural changes and the subsequent decrease in the occurrence of pulative collective modes can be attributed to the lower affinity of ATP-myosin to actin filaments, which limits the overall processivity of the myosin-II filaments. Thus the forces myosin-II filaments can exert are lower than at intermediate concentrations of ATP \cite{Cooke:1979p22092,Debold:2005p22247}. As a consequence, forced unbinding is less likely at higher ATP concentration. Accordingly, large aggregates with many crosslinkers bound cannot be disrupted while small structures with only a few crosslinking points are still ruptured and rearranged. As a consequence, material is transported from small to large and highly crosslinked structures leading to the observed cluster formation. As the material is accumulated in these clusters the connectivity between them decreases resulting in the observed low correlation.

\begin{figure}[h!]
  \centering
  \includegraphics[width=.5\textwidth]{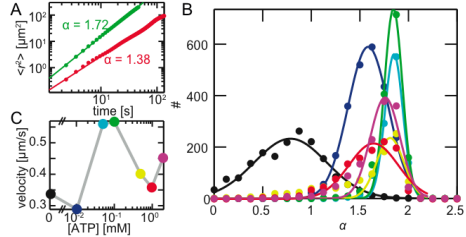}
  \caption{\textbf{Superdiffusivity in active actin networks. }%
     Mean square displacements of two individual structures at 2\,mM ATP show different but superdiffusive power law exponents (\textbf{A}). The distributions of the mean square displacement powerlaw exponents $\alpha$ show maximal superdiffusivity at intermediate ATP concentrations (\textbf{B}, 50 (cyan) to 100\,\textmu{}M (green); solid lines show gaussian fits). At low ATP concentrations (10\,\textmu{}M, blue) the low activity of myosin-II limits the superdiffusive behavior. At high ATP concentrations (0.5 (yellow), 1 (red) or 2\,mM (magenta)), myosin-II filaments are not able to exert high enough forces for constant network reorganization resulting in a lower mean superdiffusivity. Similarily, the highest mean run velocities are reached at intermediate ATP concentrations (\textbf{C}). }
  \label{fig:msd}
\end{figure}

The dynamics of the individual structures in the network is best described by the mean square displacement, which is computed for about 1500 individual points per sample. For Brownian diffusion, the mean square displacement, $\langle r^2(\tau) \rangle \propto \tau^\alpha$, is expected to increase with time $\tau$ with a power law exponent $\alpha=1$ while $0 \le \alpha<1$ or $ 1<\alpha\le2$ are indicative of sub- or superdiffusion, respectively \cite{Metzler:2000p26002,Metzler:2000p41089}. In presence of ATP, a clear superdiffusive behavior of individual clusters is found: the mean square displacements increase in time with $\alpha > 1$ (fig.~\ref{fig:msd}A) \cite{Koehler:2011}. This can be traced back to the complex alternation of runs and stalls of the individual network structures. The distribution of the powerlaw exponents depends on the ATP concentration (fig.~\ref{fig:msd}B): The passive state (i) is characterized by a subdiffusive behavior which can be attributed to the network hindering the diffusion. Already at low ATP concentrations (ii), the dynamics become superdiffusive. This superdiffusivity is even more pronounced at intermediate ATP concentrations (iii). However, at high ATP concentrations (iv) again a lower superdiffusivity is observed and the width of the distribution increases. This width is originated in the highly heterogenous network organization which predominantly occurs in this regime. Similarly to the elimination of pulsative collective modes, the lower superdiffusivity can be attributed to longer stalls (supplemental fig.~S2) due to a decrease in forced unbinding events resulting in less activity and a larger variability.

The maximal velocity of individual structures in run phases of 0.6\,\textmu{}m/s is observed for intermediate ATP concentrations while higher or lower ATP concentrations result in a decrease of the velocity in accordance with the results obtained for the pulsative collective modes or superdiffusivity. However, in all regimes the velocities in run phases are an order of magnitude lower than the velocity of myosin-II filaments observed in gliding assays at similar ATP concentrations \cite{Kron:1986p27891}. Thus, in all cases the speeds are not limited by the maximal speed of the motors, but by the presence of crosslinking bonds.

\begin{figure}[h!]
  \centering
  \includegraphics[width=.5\textwidth]{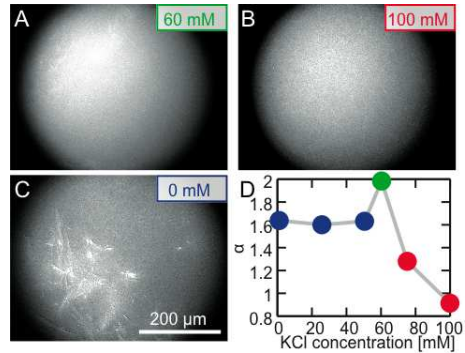}
  \caption{ \textbf{Dependence of the dynamics on the KCl concentration.} %
      Above a critical KCl concentration of 60\,mM (\textbf{A}), an active actin/fascin/myosin network cannot be distinguished from a passive actin fascin network (100\,mM, \textbf{B}), while below the critical concentration an active network is formed (\textbf{C}). The fluorescence micrographs are taken 90\,min after initiation of polymerization. The superdiffusivity as characterized by the mean mean square displacement power exponent $\alpha$ decreases above a critical KCl concentration of 60\,mM (\textbf{D}).}
  \label{fig:KCl}
\end{figure}

 Alternatively to the variation of ATP concentration, the crossbridge strength of the myosin-II filaments can be altered independently by the ionic strength \cite{Brenner:1982p7879}. An increase of the KCl concentration up to 200\,mM does not affect the myosin-II filament length \cite{Katsura:1971p7300}. We also do not observe any differences in actin/fascin network structure upon variation of the ionic strength. Indeed, exceeding a critical KCl concentration of 60\,mM prevents any dynamic reorganization within the active gel (fig.~\ref{fig:KCl}): the structure of an active actin/fascin/myosin network at more than 60\,mM KCl cannot be distinguished from a passive actin/fascin network. This critical concentration has also been observed in a gliding assay, where salt concentrations higher than 60\,mM do not support motility of actin filaments \cite{Takiguchi:1990p10397}. At this ionic strength the affinity of ATP-myosin does not allow for permanent binding of the myosin-II to the actin filaments throughout its chemomechanical cycle.

Thus, the observed network reorganization critically depends on the strength of the myosin-actin bond: Only a maximal binding strength of the myosin-II filaments to actin at low ionic strength allows exerting high enough forces to disrupt the actin/fascin crosslinking points and to reorganize the network structure. The network reorganization and the observed superdiffusivity not only rely on the competition between active and passive crosslinks but also on the respective binding strengths and affinities.

The good accessibility of all relevant system parameters within this reconstituted approach enables us to test the identified principles in a simple phenomenological simulation that is based on probabilistic interaction rules. Fascin bundles are modeled as polar rigid rods in a quasi two dimensional geometry. The bundles are propelled and crosslinked reflecting transport and passive binding processes. Both active and passive binding processes are subjected to forced unbinding events \cite{Koehler:2011}.

\begin{figure}[h!]
  \centering
  \includegraphics[width=\textwidth]{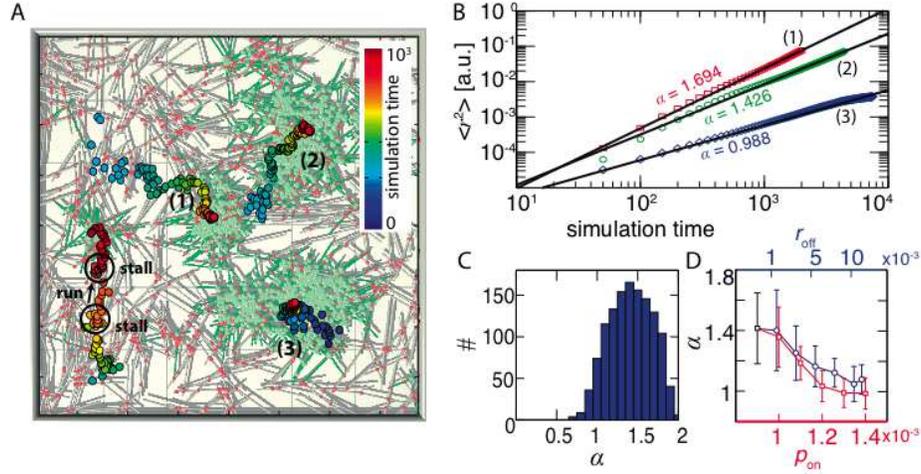}
  \caption{ \textbf{Simulation results.} (\textbf{A}) shows a simulation snapshot superimposed with the time-resolved center-off mass positions of the respective clusters. As can be seen from the center-off-mass trajectories, individual clusters move in a succession of runs and stalls and in general move less the larger they are. This is reflected in the mean-square displacements depicted in (\textbf{B}): while the large cluster shows a diffusive motion (curve 3) the other clusters move in a superdiffusive manner with mean square displacement exponents $\alpha > 1$. The heterogeneity in the system leads to a broad distribution of exponents $\alpha$ with a mean at $\alpha = 1.42\pm0.24 $ (\textbf{C}). The exact value of the mean exponent depends on the details of the motor- and crosslinker (un-)binding kinetics. Increasing the motor off-rate $r_\mathrm{off}$ leads to a decreased activity in the system, visible in the decline of the mean square displacement exponent $\alpha$ (\textbf{D}, blue curve). Likewise, the activity and hence the exponent $\alpha$ decline, if the crosslinker on-rate $p_\mathrm{on}$ is increased (\textbf{D}, red curve).
      }
  \label{fig:sim}
\end{figure}

The implementation of these basic processes already suffices to obtain a network dynamics strongly reminiscent to the experimental findings. Due to the concerted action of molecular motors, clusters are constantly transported and the trajectories they perform are characterized by a succession of runs and stalls (fig.~\ref{fig:sim}A). During run phases the velocity of the fastest clusters does not exceed one tenth of the velocity of the motor proteins; and the larger the clusters are, the lower is their susceptibility to motor induced displacements and the less they move (fig.~\ref{fig:sim}B). During stall phases clusters essentially exhibit a diffusive motion. In accordance with the experiment, runs and stalls average to a superdiffusive behavior, as can be seen in the distribution of exponents of the mean square displacement in fig.~\ref{fig:sim}C. While this is in accordance with time dependent ensemble averages \cite{Koehler:2011}, the occurence of broad distributions directly relates to the intrinsic degree of heterogeneity in the system.
Like in the experiment, the dynamics of individual clusters relies on local influencing factors like the local network connectivity, the local concentration of active crosslinks and the individual cluster size.

The exact values of $\alpha$ depend on the details of the actin-fascin and actin-myosin interaction. In the simulation, this is governed by the motor and crosslinker on- and off-rates. An increase in the motor off-rate $r_\mathrm{off}$ lowers the crossbridge strength and corresponds to the addition of KCl in the experiment. Similar to the experiment the increase of the motor off-rate hinders the reorganization in the system and individual structures, single rods or entire clusters move less. This is reflected in a reduced exponent of the mean square displacement, which successively decreases with increasing motor off-rate (fig.~\ref{fig:sim}D). If the motor activity drops below a critical value, the dynamics becomes subdiffusive. In this regime the motor proteins are not strong enough to induce any significant network reorganization and the system essentially consists of one single cluster that only shows diffusive motion. A similar behavior is found if the on-rate of the crosslinking proteins $p_\mathrm{on}$ is increased. The higher the on-rate, the more the cluster dynamics is shifted to stall phases and the exponent $\alpha$ decreases concomitantly (fig.~\ref{fig:sim}D).

Importantly, the simulation cannot retrieve the pulsative collective run and stall phases observed in the experiment, where remote structures move in a coordinated manner (fig.~\ref{fig:traje}). This indicates that the experimentally observed dynamics not only depend on the identified microscopic interactions but are modulated by a weak but far reaching connectivity within the system. While the incorporation of active transport, crosslinking and forced unbinding suffices to explain the superdiffusive transport properties, the simulation cannot account for long range force percolation as the force balance is only local with a typical range of a rodlength.

\paragraph{Discussion}
A minimal model system consisting of actin filaments, crosslinking molecules and myosin-II motor filaments exhibits drastic structural rearrangements that in turn affect the dynamic properties of the system. This feedback between large scale structure formation and dynamics is set by the microscopic details of the competition between myosin-II filaments and crosslinking molecules. The  properties of the system are sensitively controlled by the binding affinity of the myosin-II filaments to actin-fascin structures that directly regulates the maximal force the myosin-II filaments can exert. If the exerted forces are sufficiently high to disrupt actin/fascin structures, the system is dominated by the motor activity. In this regime, a highly dynamic network is formed. The dynamics in this state show pulsative collective modes which can be attributed to the connectivity of the polydisperse structures. The pulsative collective modes are not only found for directly connected structures but they are propagated throughout the network. They do not lead to a macroscopic contraction but persist in a dynamic steady state. 

The balance of motor to crosslinker strength alone sets not only the degree of dynamics in the network but also their coordination and range. A high motor activity and strength is necessary to maintain the connectivity in the network which ensures not only the occurence of pulsative collective modes but also a highly superdiffusive dynamics. Lowering the binding affinity of the motor heads limits the internal force exertion to the crosslinked actin filaments. Consequently, the crosslinking molecules dominate relative to the motor strength resulting in a non-collective dynamics with lower superdiffusivity and a crossover to normal diffusion in a highly heterogeneous network. This opens up the possibility of inducing large scale reorganizations within an active gel, building local contractile elements or establishing the systems' heterogeneity by solely changing the maximal force the motor proteins can exert. 

Any large scale reorganization within active gels depends on the intricate interplay between structure formation, aggregation and dynamic transport, which are regulated and determined by the here studied competition between force generating and force bearing structures.
For large scale reorganization of the intracellular cytoskeleton active transport of the cytoskeletal building blocks, their aggregation to higher order structure are essential for ensuing force generating capabilities. The dynamics of these processes will depend on the here studied interaction and competition between the force generating and force bearing structures. The combination of simplified \emph{in vitro} model systems with simulations based on simple interaction rules opens up the perspective of addressing the governing principles of the cytoskeletal structure formation processes.

\paragraph{Materials and Methods}

\noindent \emph{Protein purification}
Myosin  \cite{Margossian:1982p24829} and G-actin \cite{Spudich:1971p24798,MacLeanFletcher:1980p24822} are extracted from rabbit skeletal muscle (rabbit meat was obtained from \emph{Hasenhof Weh}, Moorenweis, Germany). To fluorescently label actin, actin is dialyzed against 50\,mM boric acid (pH~8.0), 0.2\,mM CaCl$_2$, 0.2\,mM ATP and polymerized by addition of 2\,mM MgCl$_2$ and 100\,mM KCl. F-actin is incubated with Alexa Fluor 555 succinimidylester (Invitrogen) at a molar ratio of 1:1 at room temperature for 1\,h and centrifuged at 100\,000\,g at 4\,$^{\circ}$C for 2\,h. Labelled F-actin is depolymerized by dialysis against G-buffer (2\,mM Tris/HCl (pH~8.0), 0.2\,mM CaCl$_2$, 0.2\,mM ATP, 0.2\,mM DTT) at 4\,$^{\circ}$C. This yields a 25\,\% degree of labelling.
Recombinant human fascin is purified from \emph{E.coli} BL21-CodonPlus-RP and stored at -80\,$^{\circ}$C in 2\,mM Tris/HCl (pH~7.4), 150\,mM KCl at 64\,\textmu{}M \cite{Vignjevic:2003p24834}.

\emph{Fluorescence imaging}
Actin is polymerized at 1\,\textmu{}M by adding one-tenth of the sample volume of a 10-fold concentrated polymerization buffer (100 \,mM imidazole, 2\,mM CaCl$_2$, 30\,mM MgCl$_2$), 0.1\,\textmu{}M myosin-II, and 1\,\textmu{}M fascin, 20\,mM creatine phosphate and 0.1\,mg/mL creatine phospho kinase (Sigma) for ATP regeneration, 3\,mg/mL casein, ATP and KCl at indicated concentrations. In this polymerization buffer, myosin readily polymerizes into filaments with a mean length of 0.6\,\textmu{}m independent of the KCl concentration. Samples are enclosed to hermetically sealed chambers to eliminate any drift in the network \cite{Koehler:2011}.
\newline
\noindent All data are acquired on a Zeiss Axiovert 200 inverted microscope with either a 10\,x (NA 0.2) long distance objective or a 40\,x (NA 1.3) oil immersion objective. Images are captured at 0.84\,frames/s with a charge-coupled device camera (Orca ER, Hamamatsu) attached to the microscope via a 0.4\,x camera mount.

\noindent\emph{Image processing}
For tracing individual actin structures, images are background subtracted in ImageJ. To identify individual points in the network, an intensity threshold value is applied in ImageJ to generate a binary image. Individual structures are identified as more than 10 connected bright pixels. The structures are traced over time using the IDL tracking algorithm by John C.~Crocker \cite{Crocker:1996p44059} for the intensity weighted centroid positions using Matlab R2008b (The MathWorks, Inc.). To minimize tracking artefacts, the trajectories are subjected to a gliding average over 4 frames. Subsequently, the trajectories are divided into "runs" and "stalls":
Runs are identified as movements between two frames with velocities larger than 0.36\,\textmu{}m/s and a change in direction smaller than 45\,$^\circ$ and the lengths of continuous runs over 2 successive frames are calculated. Run velocities are calculated as mean velocity of a single run and mean values are obtained by fitting Lorenz distributions. 
Mean square displacements are calculated for entire trajectories and the first 10\,\% of the resulting mean square displacements are used. The power exponent $\alpha$ is fitted logarithmically to mean square displacements longer than 5 time points. Average values of $\alpha$ are obtained by fitting a Gaussian to the distribution of the power exponent to account for the heterogeneity of the networks. 
\newline
\noindent To determine phases of correlated movement, the velocity cross-correlation function of moving structures $I_v(\mathbf{r})= \big[\langle v(\mathbf{x}+\mathbf{r})v(\mathbf{x})\rangle_{\mathbf{x}} - \langle v\rangle_{\mathbf{x}}^2\big]/ \big( \langle v^2\rangle_{\mathbf{x}}- \langle v\rangle_{\mathbf{x}}^2 \big)$ is evaluated. The degree of correlation is proportional to the average of the squared correlation function $\langle I^2_v(\mathbf{r})\rangle_r$ (fig.~S1). The proportion of time points showing collective modes is evaluated by introducing a correlation cutoff $k$: Averaged squared correlation functions $\langle I^2_v(\mathbf{r})\rangle_r$ above $k$ are defined as collective modes; values below $k$ are defined to relate to an uncorrelated movement. For the data displayed in fig.~\ref{fig:traje} $k$ was set to 0.15. The variation of $k$ only affects the absoulte values while the relative values and the trend are not affected.

\noindent\emph{Simulations}
The experimental system consists of interconnected polar fascin bundles that are actively transported and set under stress by myosin-II filaments \cite{Koehler:2011}. The microscopic processes that lead to the observed self organization, arise through the interplay of crosslinking events and active transport. More specifically, the competition between molecular motors and crosslinking proteins gives rise to the active movement, (re-)binding and forced unbinding.
These microscopic interactions are the basis for numerical simulations.
In a minimal approach fascin bundles are modelled as monodisperse and polar rigid rods in a two dimensional geometry.
If two filaments overlap active or passive binding events occur based on probabilistic interaction rules governed by the motor and crosslniker on rates $r_\mathrm{on}$ and $p_\mathrm{on}$. In a similar manner unbinding processes are calculated with the motor and crosslinker off-rates $r_\mathrm{off}$ and $p_\mathrm{off}$. The displacements that arise due to the action of molecular motors are calculated based on generic velocity models similar to those presented in Ref.~\cite{Liverpool:2003p30078}. Both active and passive bindings are subjected to forced unbinding events: Conceptually the stress increase leads to rupture events within the cluster whereby especially weakly crosslinked structures with a low actin and/or crosslinker density are prone to rupture. In a simplified picture this can be modelled by increasing the off-rate with the number of motor proteins per individual cluster. If not indicated otherwise, all simulations were performed with a density $\rho=14/L^2$, a motor on-rate $r_\mathrm{on}=0.1$, a motor off-rate $r_\mathrm{off}=9\cdot10^{-4}$, a crosslinker on-rate $p_\mathrm{on}=9\cdot10^{-4}$ and a crosslinker off-rate $p_\mathrm{off}=1\cdot10^{-4}$.

\paragraph{Acknowledgments}
We thank the DFG in the framework of the SFB 863, the German Excellence Initiative 'International Graduate School for Science and Engineering' and 'Nanosystems Initiative Munich' for financial support. V.S. acknowledges support from the Elite Network of Bavaria by the graduate programme CompInt.


\clearpage

\end{document}